\documentclass[pre,twocolumn]{revtex4}

\usepackage{amsmath}
\usepackage{amssymb}
\usepackage{graphicx}

\begin{document}

\title{Econophysics of interest rates and the role of monetary policy}


\author{Daniel O. Cajueiro$^\star$ and Benjamin M. Tabak$^{\star\star}$}

\affiliation{$^\star$ Universidade Cat\'{o}lica de Bras\'{i}lia --
Graduate Program in Economics. \\SGAN 916, M\'odulo B -- Asa
Norte. DF 70790-160 Brazil. \\ $^{\star\star}$ Banco Central do
Brasil\\ SBS Quadra 3, Bloco B, 9 andar. DF 70074-900}

\begin{abstract}
This paper presents empirical evidence using recently developed
techniques in econophysics suggesting that the degree of
long-range dependence in interest rates depends on the conduct of
monetary policy. We study the term structure of interest rates for
the US and find evidence that global Hurst exponents change
dramatically according to Chairman Tenure in the Federal Reserve
Board and also with changes in the conduct of monetary policy. In
the period from 1960's until the monetarist experiment in the
beginning of the 1980's interest rates had a significant
long-range dependence behavior. However, in the recent period, in
the second part of the Volcker tenure and in the Greenspan tenure,
interest rates do not present long-range dependence behavior.
These empirical findings cast some light on the origins of
long-range dependence behavior in financial assets.
\end{abstract}

\maketitle
%
%
%

\section{Introduction}
\label{int}

In the past decades the US economy has experienced low inflation
and little variation in real activity if compared to the 1970's.
These improvements have been largely attributed to a change in the
way the Federal Reserve conducts monetary policy. A number of
research papers have suggested that a structural break in the
conduct of monetary policy has occurred since Paul Volcker became
chairman of the Federal Reserve in August 1979 (see Clarida et
al., 2000).

However, there is little consensus as whether a change in the
conduct of monetary policy has indeed occurred and if it has what
would be the dates of these changes (Boivin, 2005).

This paper adds to the debate on monetary policy by studying
changes in persistence in interest rates for different maturities
for the US. We investigate 1, 3, 5 and 10 year maturity interest
rates and present overwhelming evidence that a structural break
has occurred in the dynamics of these interest rates. We employ
methods recently developed in statistical physics and show that
interest rates' persistence has decreased substantially in the
post-1982 period, while there is evidence of strong long-range
dependence in the pre-1982 period. Therefore, the evidence in this
paper is in line with the reasoning that a structural break has
occurred in the conduct of monetary policy in the early 1980's.

This paper proceeds as follows. In section 2, a brief review of
the literature is presented. In section 3, the methodology to
estimate generalized Hurst exponents is reviewed. In section 4,
the data used in this work is described. Section 5 presents the
empirical results. Finally, in section 6, this paper is concluded.

\section{Brief Literature Review}
\label{sec:BRL}

Researchers have documented a substantial change in macroeconomic
variables for the US in the past decades. From the late 1960s
through the early 1980s, the United States economy experienced
high and volatile inflation along with several severe recessions.
Since the early 1980s, however, inflation has remained steadily
low, while output growth has been relatively stable.

When Arthur Burns became chairman in February 1970, the inflation
rate had reached nearly 6 percent, yet almost immediately,
monetary policy became more accommodating in that the Federal Open
Market Committee (FOMC) lowered the targeted federal funds rate
and money growth exploded. Over the period 1965 through the end of
the 1970s, monetary policy earned the appellation "stop-go" from
the FOMC's alternate concentration on reducing inflation and
stimulating economic activity.

The initial period of Volcker's tenure as chairman, August 1979
through 1982, can be thought of as the final stop phase of the
preceding stop-go period. In that period the  Federal Reserve
announced a change in operating procedures from partially
targeting interest rates to targeting nonborrowed reserves. This
period, from October 1979 to August 1982, is known as the
monetarist experiment.

In 1982 the Volcker Federal Reserve began targeting the fed-funds
rate. From the mid-1980's onward, monetary policy consistently
responded strongly to inflation and weakly to real-activity. This
interest-rate targeting continued in the Greenspan administration.

Romer and Romer (2003) study why monetary policy has been so much
successful under some Federal Reserve chairmen than others. The
authors argue that the key determinants of policy success have
been policymakers views about the economy and limitations of
monetary policy.

Several papers have documented that a change in monetary policy
has occurred in the early 1980's. Clarida {\it et al.} (2000)
provide empirical evidence of important changes in the U.S.
conduct of monetary policy over the last forty years. In
particular they find that while monetary policy accommodated
inflation in the 1970's, this drastically and suddenly changed
with the appointment of Volcker in 1979. They emphasize that the
pre-Volcker conduct of monetary policy did not satisfy the
so-called Taylor principle, so that a given increase in inflation
was typically associated with a smaller increase in the nominal
interest rate. The authors show that in the Volcker-Greenspan era
the Federal Reserve adopted a proactive stance toward controlling
inflation. Duffy and Engle-Warnick (2004) study changes in
monetary policy over the 1995-2003 period and find evidence o
three structural breaks, one of them in the beginning of the
Volcker's Federal Reserve chairmanship\footnote{Sims and Zha
(2004) argue that the only period since 1950 with a noticeably
different monetary policy is the monetarist experiment between
1979 and 1982, in which the Federal Reserve targeted monetary
aggregates. They do not find any differences in the monetary
policy behavior in the 1970's and 1980's.}.

The literature cited above has discussed whether and when
structural breaks have occurred in the conduct of monetary policy.
The overall conclusion is that changes have occurred in the
dynamics of inflation and real activity. Therefore, we will study
the dynamics of interest rates (long memory) and test whether
changes have occurred in the dynamics of these variables
evaluating different time periods, according to hypothesized
changes in monetary policy and Federal Reserve chairman.

The first to consider the existence of long memory in interest
rates seems to be Backus and Zin (1993) who find evidence of
long-memory in the 3-month zero-coupon rate for the US, and that
allowing for long memory in the short interest rate improves the
fitted mean and volatility yield curves. The authors have
suggested that the sources of the long-memory property of the
short-term interest rate may be derived from a fractionally
integrated dynamic for inflation and/or the money growth rate.

Since then, others have supported Backus and Zin (1993) results.
For example, Tsay (2000) has showed that the ex post real interest
rate for the US possesses long memory. The author employed unit
root tests due to Kwiatkowski {\it et al.} (1992) and Phillips and
Perron (1988). For most of the samples the authors analyzes the
rejection of both hypothesis suggest that these process are
neither an I(1) or I(0) process. Therefore, an ARFIMA model was
estimated and empirical evidence suggested that ex post real
interest rates could be well described by an ARFIMA model with
long memory. Other evidences are provided by Barkoulas and Baum
(1998) employing spectral regression and Gaussian semiparametric
methods to the Euroyen deposit rates and Euroyen term premium and
McCarthy {\it et al}. (2004) applying wavelets to a large class of
US debt instruments. In particular,the authors in Cajueiro and
Tabak (2005a) have found an interest evidence of this phenomenon
studying the long-range dependence behavior of the term structure
of interest rates in Japan where the predictability in the term
structure of interest rates increases with maturity. In their
paper this phenomenon is explained by the nonnegative constraint
in the interest rate.

The contribution of this paper is that we study long memory
properties for interest rates for different time periods and check
whether a structural break has occurred, which is suggestive of
changes in monetary policy.

\section{Measures of Long-Range Dependence}
\label{sec:lrd}

Several methods have been introduced to take the long-range
dependence phenomenon into account\footnote{A survey of these
methods may be found in Taqqu {\it et al.} (1999).}. This
literature can be actually divided in two different strands: (1)
an approach whose focus is to determine the parameter known as the
Hurst exponent or a parameter related to it (see, for example
Geweke and Porter-Hudak (1983), Hosking (1981), Hurst (1951),
Robinson (1995) and Cajueiro and Tabak (2005b)) and (2) an
approach that aims at developing statistics to test, through a
hypothesis test, the presence of long-range dependence (see, for
example, Giraitis {\it et al.} (2003), Lee and Schmidt (1996) and
Lo (1991)).

In this paper, our measure of long range dependence is the
Generalized Hurst exponent introduced in Barabasi and Vicsek
(1991) and considered recently by Di Matteo {\it et al.} (2005) to
study the degree of development of financial markets. The
generalized Hurst exponent is a generalization of the approach
proposed by Hurst. The authors suggests analyzing the
\textit{q}-order moments of the distribution of increments, which
seems to be a good characterization of the statistical evolution
of a stochastic variable $X(t)$,

\begin{equation}K_{q}(\tau)=\frac{\langle|X(t+\tau)-X(t)|^{q}\rangle}{\langle|X(t)|^{q}\rangle},\end{equation}

where the time-interval $\tau$ can vary\footnote{For $q=2$, the
$K_{q}(\tau)$ is proportional to the autocorrelation function
$\rho(\tau)=\langle X(t+\tau)X(t)\rangle$.}. The generalized Hurst
exponent can be defined from the scaling behavior of
$K_{q}(\tau)$, which can be assumed to follow the relation

\begin{equation}K_{q}(\tau)\sim(\frac{\tau}{\nu})^{qH(q)}.\end{equation}

\section{Data}
\label{sec:data}

The data is sampled daily, beginning on January 2, 1962 and ending
on February 4, 2005. The full sample has 10755 observations,
collected from the Federal Reserve System. We study the 1,3 ,5 and
10-years to maturity interest rates, which are constant maturity
treasury rates.

We test for long-range dependence in log interest rates for
different time periods. We split the sample according to monetary
policy and also to Federal Reserve tenure. Table 1 presents the
tenure period for each chairman. We do not study the Miller
administration because it was too short.

\begin{table}
\begin{tabular}{cc}

\hline

       Federal Reserve Chairman  &  Period   \\
\hline

W. Martin & Apr. 1951 - Jan. 1970\\
A. Burns & Feb. 1970 - Jan. 1978\\
G. Miller & Mar. 1978 - Aug. 1979\\
P. Volcker & Aug. 1979 - Aug. 1987\\
A. Greenspan & Aug. 1987 - Feb. 2006\\

\hline
\end{tabular}
\begin{flushleft}{This table presents the tenure of each Chairman of the Federal Reserve since the 1950's.}\end{flushleft}
\end{table}

\section{Empirical Results}
\label{sec:data}

Recent research has documented that a change may have occurred in
the way monetary policy has been conducted in the US in the past
decades (see Clarida {\it et al.}, 2000, and Boivin, 2004).
Therefore, we study the behavior of interest rates for different
maturities and compare generalized Hurst exponents for a variety
of time periods.

Table 2 presents generalized Hurst exponents for different time
periods. Panel A presents estimates according to Federal Reserve
chairman. Hurst exponents are decreasing with maturity, which
suggests that short-term interest rates are more predictable than
long-term interest rates. It is striking that these Hurst
exponents are close to 0.5 for the Greenspan era for all
maturities, and are very high for the Burns era (above 0.62 for
all maturities).

We would also like to test whether there is an influence of the
monetarist experiment conducted in the beginning of the Volcker
administration. Panel B shows results dividing the sample in a
different way. We see that interest rates were quite persistent in
the monetarist experiment in the beginning of the Volcker
administration. However, they converge to values similar to the
ones seen in the Greenspan administration afterwards.

\begin{table}
\begin{tabular}{ccccc}

\hline

    &   y1  &   y3  &   y5  &   y10 \\
\hline
Panel A: Federal Reserve Chairman & & & & \\
Martin  &    0.64   &    0.59   &    0.59   &    0.59   \\

Burns   &    0.64   &    0.63   &    0.62   &    0.62   \\

Volcker &    0.58   &    0.58   &    0.58   &    0.56   \\

Greenspan   &    0.50   &    0.50   &    0.50   &    0.49   \\

Panel B: Monetary Policy &   &   &   &   \\

Pre 1979    &    0.63   &    0.61   &    0.61   &    0.61   \\

Post 1979   &    0.53   &    0.52   &    0.52   &    0.51   \\

Monetarist Experiment   &    0.60   &    0.59   &    0.59   &    0.57   \\

Post 1982   &    0.50   &    0.50   &    0.51   &    0.50   \\
\hline
\end{tabular}
\begin{flushleft}{This table presents generalized Hurst exponents for 1,3,5 and 10-year interest rates for different time periods.  }\end{flushleft}
\end{table}

The empirical results obtained suggest that the dynamics of
interest rates has changed substantially in the past decades.
Long-range dependence seems to be strong in the pre-1982 period,
while this evidence practically disappears in the recent period
(post-1982), coinciding with substantial changes in the conduct of
monetary policy.

\section{Conclusions}
\label{sec:conc}

Testing for long-range dependence in asset prices has been subject
of intense investigation in the financial literature. There are
many implications for portfolio and risk management. For example,
traditional option pricing models should be modified to
incorporate long-range dependence features in asset prices and
volatility. Furthermore, if the long-range dependence parameters
change over time, then the time series that are being studied
possess more information than is given by monofractal models.
Therefore, studies that focus on how and why long-range parameters
change over time may be particularly useful as they can be used to
determine structural breaks or shifts in these time series.

This paper offers a fresh look at the properties of interest rates
for the US. The empirical evidence suggests that interest rates
had strong long memory in the pre-Volcker administration and that
after 1982 this evidence has disappeared. These results suggest a
structural break in the dynamics of interest rates. They also
imply that careful should be taken when studying long time series
as the parameters that characterize them may change over time,
which is evidence of multifractality.

It is important to notice that our sample period includes
important changes in the macroeconomic environment, as exchange
rates become flexible in the mid 1970's and early 1980's.
Therefore, in a fixed exchange rate framework shocks to the
economy must be absorbed mainly by movements in interest rates,
which implies in more persistent interest rates' dynamic. However,
in flexible exchange rate regimes policy makers have more degrees
of freedom to absorb shocks into the economy, as exchange rates
may absorb partially such shocks.

\section{Acknowledgements} The authors thank participants of the AFPA5 2006 for helpful suggestions. Benjamin M. Tabak gratefully acknowledges financial support from CNPQ foundation. The opinions expressed in this paper are those of the authors and do not necessarily reflect those of the Banco Central do Brasil.


\begin{thebibliography}{00}


\bibitem{baczin93}\textsc{Backus,D, Zin, S.} Long memory inflation uncertainty: evidence of term structure of
interest rate. {\it Journal of Money, Credit and Banking}, {\bf
25}, 687-700, 1993.

\bibitem{bar} \textsc{Barabasi, A.L., and Vicsek, T.} Multifractality of self-affine fractals.
{\it Physical Review A}, {\bf 44}, 2730-2733, 1991.

\bibitem{barbau98} \textsc{Barkoulas, J. T. and Baum, C. F.} Fractional dynamics in Japanese financial time
series. {\it Pacific-Basin Finance Journal}, {\bf 6}, 115-124,
1998.

\bibitem{boi3}\textsc{Boivin, J.} Has US Monetary Policy Changed? Evidence from Drifting
Coefficients and Real-Time Data. {\it NBER Working Paper}, 2005.

\bibitem{cajtab05}\textsc{Cajueiro, D. O. and Tabak, B. M.} { The long-range dependence behavior of the term structure of interest rates in Japan}. {\it Physica A}, {\bf 350}, 418-426, 2005a.

\bibitem{cajtab05b}\textsc{Cajueiro, D. O. and Tabak, B. M.} {The rescaled variance statistic and the determination of the Hurst's exponent}. {\it Mathematics and Computers in Simulation}, 2005b.

\bibitem{cajtab05b}\textsc{Clarida, R., Galí, J., and Gertler, M.} {Monetary Policy Rule and
Macroeconomic Stability: Evidence and Some Theory}. {Quarterly
Journal of Economics}, {\bf 115}, 147-180, 2000.

\bibitem{di}\textsc{Di Matteo, T., Aste, M., and Dacorogna, M.} Long-term memories of developed and emerging markets: Using the scaling analysis to characterize their stage of development.
{\it Journal of Banking and Finance}, {\bf 29}, 827-851, 2005.



\bibitem{duf}\textsc{Duffy, J., and Engle-Warnick, J.}   Multiple Regimes in U.S. Monetary Policy?
A Nonparametric Approach. {\it Forthcoming in the Journal of
Money, Credit and Banking}, 2004.

\bibitem{gew83}\textsc{Geweke, J. and Porter-Hudak, S.} The estimation and application of long memory time series models. {\it Journal of Time Series Analysis},
{\bf 4}, 221-238, 1983.

\bibitem{girkok03}\textsc{Giraitis, L., Kokoszka, P., Leipus, R. and Teyssi\`{e}re, G.} Rescaled variance and related tests for long memory in volatility and levels.  {\it Journal of Econometrics},
{\bf 112}, 265-294, 2003.

\bibitem{hos81}\textsc{Hosking, J. R. M.}Fractional Differencing. {\it Biometrika},
{\bf 68}, 165-176, 1981.

\bibitem{hur51}\textsc{Hurst, E.} Long term storage capacity of reservoirs. {\it Transactions on American Society of Civil Engineering}, {\bf
116}, 770-808, 1951.

\bibitem{kw92}\textsc{Kwiatkowski, D., Phillips, P.C.B., Schmidt, P., Shin, Y.,}
Testing the null hypothesis of stationarity against the
alternative of a unit root. {Journal of Econometrics}, {\bf 54},
159-178, 1992.

\bibitem{lee96}\textsc{Lee, D. and Schimidt, P.} On the power of KPSS test of stationarity against fractionally-integrated alternatives.{\it Journal of Econometrics}, {\bf
73}, 285-302, 1996.

\bibitem{lo91}\textsc{Lo, A. W.} Long-term memory in stock market prices. {\it Econometrica},
{\bf 59}, 1279-1313, 1991.

\bibitem{mac04}\textsc{McCarthy, J., DiSario, R., Saraoglu, H., Li, H.}
Tests of long-range dependence in interest rates using wavelets.
{\it The Quarterly Review of Economics and Finance}, {\bf 44},
180-189, 2004.

\bibitem{phiper88}\textsc{Phillips, P. C. B. and Perron, P.} Testing for a unit root test in time series regression. {\it Biometrika},
{\bf 75}, 335-346, 1988.

\bibitem{rob95}\textsc{Robinson, P. M.} Gaussian semiparametric estimation of long-range dependence. {\it The Annals of Statistics},
{\bf 23}, 1630-1661, 1995.

\bibitem{ts0}\textsc{Romer, C. and Romer, D.} Choosing the Federal Reserve Chair: lessons from history.
{\it NBER Working Paper},  2003.

\bibitem{t0}\textsc{Sims, C. and Zha, T.} Where there Regime Switches in US
Monetary Policy? {\it Working Paper Princeton University}, 2004.

\bibitem{taqtev99}\textsc{Taqqu, M. S., Teverovsky, V. and Willinger, W.} Estimators for long-range dependence: an empirical study.{\it Fractals},
{\bf 3}, 785-798, 1999.

\bibitem{tsa00}\textsc{Tsay, W. J.} Long memory story of the real interest rate. {\it Economics Letters}, {\bf
 67 }, 325– 330, 2000.

\bibitem{wiltaq99}\textsc{Willinger, W., Taqqu, M. S. and Teverovsky, V.} Stock market prices and long-range dependence. {\it Finance and Stochastics},
{\bf 3}, 1-13, 1999.


\end{thebibliography}
\end{document}